\documentclass[aps,prb,twocolumn,showpacs,floatfix,preprintnumbers,amsmath,amssymb]{revtex4-2}
\usepackage{graphicx}
\usepackage{dcolumn}
\usepackage{bm}
\usepackage{physics}
\usepackage{amsmath}
\usepackage{amssymb}

\usepackage{xcolor}
\usepackage{mathtools}

\usepackage[capitalize]{cleveref}

\begin{document}
\title{Dynamical transitions in aperiodically kicked tight-binding models}

\author{Vikram Ravindranath}
\email{vravindranath@gradcenter.cuny.edu}
\affiliation{Department of Physics and Astronomy, CUNY College of Staten Island, Staten Island, NY 10314,\\ 
Physics Program, The Graduate Center, CUNY, New York, NY 10016.}
\author{M.\ S.\ Santhanam}
\email{santh@iiserpune.ac.in}
\affiliation{Indian Institute of Science Education and Research, Dr. Homi Bhabha Road, Pune 411 008, India.}

\begin{abstract}
If a localized quantum state in a tight-binding model with structural aperiodicity is subject to noisy evolution, then it is generally expected to result in diffusion and delocalization. In this work, it is shown that the localized phase of the kicked Aubry-Andr\'{e}-Harper (AAH) model is robust to the effects of noisy evolution, for long times, provided that some kick is delivered once every time period. However, if strong noisy perturbations are applied by randomly missing kicks, a sharp dynamical transition from a ballistic growth phase at initial times to a diffusive growth phase for longer times is observed. Such sharp transitions are seen even in translationally invariant models. These transitions are related to the existence of flat bands, and using a 2-band model we obtain analytical support for these observations. The diffusive evolution at long times has a mechanism similar to that of a random walk. The time scale at which the sharp transition takes place is related to the characteristics of noise. Remarkably, the wavepacket evolution scales with the noise parameters. Further, using kick sequence modulated by a \lq{}coin toss\rq{}, it is argued that the correlations in the noise are crucial to the observed sharp transitions.
\end{abstract}
	
\maketitle

\section{Introduction}
\label{intro}
   It is by now well established through theoretical studies and experiments that periodic forcing imparted to quantum systems can lead to novel states of matter that did not exist in its time-independent counterpart \cite{floquet-engg-1,floquet-engg-2,floquet-engg-3,mbl-floquet}. For the periodically driven Hamiltonian systems, Floquet states represent the natural generalization of the stationary eigenstates for time-independent systems. These states are interesting objects of study because they can exhibit more complex dynamics and allow more control through external driving in comparison to their static counterparts. Many phenomena in static systems such as the topology of band structures \cite{PhysRevB.95.035136} and transport properties \cite{PhysRevA.98.013635,PhysRevLett.67.516} can be tuned by an appropriate driving mechanism. A class of such techniques now known as Floquet engineering \cite{floquet-engg-1,floquet-rev-2,PhysRevLett.123.216803} attempt to create novel Floquet states with desired topological properties \cite{floquet-prx,PhysRevLett.123.266803} by designing an effective time-independent Hamiltonian for the driven systems. In the last decade, experiments have implemented a variety of approaches based on such Floquet manipulations -- effective Hofstadter Hamiltonian using cold atoms in optical lattices \cite{PhysRevLett.111.185301}, control over direction and interference of phonon flow in 2D array of trapped ions \cite{PhysRevLett.123.213605}, the implementation of Haldane model in ultracold fermions to realize topological insulators \cite{haldane-model} and control over heating due to periodic drive in pre-thermal phase of Bose-Hubbard system \cite{BH-control}. Recently, principles of Floquet engineering have been extended to quantum dissipative systems as well thus providing a handle to understand non-equilibrium steady states \cite{floquet-engg-ness}.

In this broader context, of particular interest in a variety of condensed matter systems is the observation of quantum localization of eigenstates in time-independent systems and of Floquet states in time-dependent systems \cite{abrahams2010, MBL-1}. Quantum kicked rotor, representing the dynamics of a periodically kicked pendulum, is a well studied example of a classically non-integrable system. The classical limit of kicked rotor, for strong kick strengths, displays chaotic dynamics and energy diffusion \cite{reichl}. The corresponding quantum system  exhibits stark differences from classical regime, and localization of all its Floquet states is a notable feature amongst them \cite{reichl}. The quantum kicked rotor has been mapped to the Anderson tight binding model for a particle in a crystalline lattice with random on-site potentials \cite{PRL.49.509}. These connections have been experimentally explored using cold atoms in optical lattices over the last three decades \cite{PRL.75.4598}. Thus, far from being theoretical constructs, tight-binding models have become popular partly owing to their ease of experimental realisation in optical lattices \cite{RevModPhys.89.011004}. 

Floquet systems are characterised by periodic driving with periodicity $T$ such that drive term $f(t)$ satisfies $f(t+T) = f(t)$. In typical experimental situations, there are bound to be imperfections and this provides a motivation to consider noise in $T$ and other parameters related to the drive term. More importantly, there is a strong possibility of novel effects arising due to presence of aperiodicity in Floquet systems.  For instance, all the Floquet states of the quantum kicked rotor in one and two dimensions are localized, and do not display a localization to delocalization transition. However, noisy kick sequences can induce such a transition. If the kick period is perturbed by stationary noise, dynamical localization is destroyed through an exponentially decaying decoherence process resulting in a delocalized phase. This aspect has been extensively studied in kicked rotors \cite{PhysRevLett.81.1203,PhysRevLett.53.2187,PhysRevA.44.2292}. Interestingly, recent theoretical proposals \cite{PRA-levy} and experimental realizations show that the localization to delocalization transition can be non-exponential upon introducing nonstationary noise in the kicking sequence \cite{PRL-santh}. In general, without noisy kicks, the localization to delocalization transition is absent in the one-dimensional kicked rotor.

Though the localization-delocalization transition is absent in kicked rotors (and also in Anderson model) of one and two dimensions, it is known that quasi-periodic lattice models such as the Aubry-Andr\'{e} model do display such a transition even in one dimension \cite{AAmodel-MBL-review}. Depending on the presence or absence of interactions and the parametric regime under consideration, particular effects of {\sl periodically} driving the Aubry-Andr\'{e} system vary from preserving the localization transition \cite{bordia2017periodically}, inducing delocalization \cite{PhysRevE.97.010101} to the appearance of a Griffiths-like phase manifesting as slow dynamical spreading \cite{PhysRevResearch.1.032039}. Generally, the phase boundaries depend on the amplitude and frequency of the driving field \cite{PhysRevA.98.013635}. 

Then, the question arises as to how such systems with built-in structural aperiodicity react to noisy driving fields or kick sequences imparted to them. In this paper, the kicked Aubry-Andr\'{e} system, belonging to a class of aperiodic lattice models, is studied to examine the effects of noisy kicks in which noise manipulates the periodic kick sequence in three different ways. In this context, it is of interest to distinguish between milder forms of noisy kicks in which only the amplitude of kicks are modulated as opposed to the stronger forms in which entire kicks can be missed. If a tight-binding model that supports a localized regime is {\sl irregularly} kicked, we generally expect to observe a localization to delocalization transition analogous to the behaviour of the quantum kicked rotors under milder forms of noisy kicks \cite{PhysRevLett.81.1203,PRL-santh,PRA-levy}. For instance, such a scenario also unfolds in the case of Floquet topological chains treated analytically using the Floquet superoperator formalism \cite{PhysRevB.98.214301}. In this case, strictly periodic kicks preserve topologically protected end states, and noisy kicks lead to a decay of these topologically protected modes. It is also known that localized regime in the kicked Aubry-Andr\'{e} model survives even if a milder form of noise is present \cite{PRB2}. 

In contrast, in this study, the focus is on the effects of stronger forms of noisy kick sequences imparted to an initially localized state. It is shown that a dynamical transition from ballistic to diffusive growth results. The characteristics of this transition, whether smooth or abrupt, and the timescale at which this transition takes place depend on the noise properties. Remarkably, by tuning the characteristics of noise, the timescale for the transition from ballistic to diffusive can be tuned as well. It is shown that the wavepacket growth quantified by its second moment scales with a noise parameter. Rest of the article is structured as follows -- in sections \ref{sec2} and \ref{sec3}, we describe the AAH model and noisy driving protocols, in sections \ref{sec4} and \ref{sec5} the simulation results are presented, and in sections \ref{sec6} and \ref{sec7} the results are extended to translationally invariant models and analytical support is provided for the main features of the results. Section \ref{sec8} provides a brief summary of the results.

\section{Dynamical Properties}
\label{sec2}
\subsection{Model}
The Hamiltonian of the model considered in this work is given by
\begin{equation}
H = H_0 + V\sum_{n}\delta(t-T_n)\sum_{i=1}^{L}V_i ~c_i^\dag c_i,
\label{model}
\end{equation}
where the integrable part of the Hamiltonian representing the kinetic energy term is
\begin{equation}
H_0  = -J\sum_{i=1}^{L} \left(c_i^\dag c_{i+1} + c_{i+1}^\dag c_i\right).
\label{model-intpart}
\end{equation}
In this, $c_i^{\dagger}$ and $c_i$ are the creation and annihilation operators at $i$-th site. This system has $L$ lattice sites with periodic boundary conditions applied to it. The onsite potential is $V_i \equiv \cos(2\pi\alpha i)$ and $\alpha$ is a real parameter that modulates the aperiodicity of the lattice. The temporal dependence in the form of Dirac delta function in Eq. \ref{model} ensures that the onsite potential kicks-in whenever $T_n=n \tau$, where $n$ is an integer and $\tau$ is the kicking period. The parameter $V$ represents the amplitude of periodic kicks and $J$ is the amplitude for the strength of nearest-neghbour hopping.

Without the train of $\delta$-kicks and if $\alpha$ is an irrational number, then Eq.\ \ref{model} is the well-known Aubry-Andr\'{e} model \cite{aubry} for electron transport in a lattice with structural disorder. Extensive investigation of this model \cite{10.2307/121066} has shown that if $\alpha$ is a strongly incommensurable number such as the Golden ratio, then all the eigenstates display a transition from extended to localized states at the critical point $V/J = 2$. The transition to localized states has been experimentally observed in the test-bed of non-interacting ultracold atoms in quasiperiodic optical lattices \cite{roati2008anderson}, as well as in the interacting case \cite{Schreiber842}. Numerical evidence too points to the existence of a {\sl many-body} localization transition in the interacting Aubry-Andr\'{e} model \cite{PhysRevB.87.134202}. The corresponding asymptotic dynamics of an initially localized wavepacket with width $\sigma$ is ballistic, and the wavepacket width grows as $\sigma^2_t \sim t^2$, for $V/J < 2$. However, if  $V/J > 2$ the spread of the wavepacket is completely suppressed \cite{S0217979292000153}, $\sigma^2_t \sim t^0$. This transition is accompanied by multifractality in the wavefunction \cite{S0217979292000153}, diffusive dynamics \cite{10.1143/JPSJ.57.1365} and a fractal spectrum at the transition point \cite{PhysRevLett.66.1651,PhysRevB.14.2239}. The dynamics can even be strongly hyperballistic when a quasiperiodic section is embedded in a tight-binding lattice \cite{PhysRevLett.108.070603}.

If the periodic kicks are applied at times $T_n=n \tau$, ($n \in \mathbb{Z}^+$), the critical point for localization transition in the limit of high-frequency of kicks is found to be $\frac{V}{JT} \sim 2$ \cite{PRB1}. However, if this critical point is approached from the delocalized phase, the system exhibits a range of growth rates; {\it i.e.,} $\sigma^2_t \sim t^\gamma, 1 \leq \gamma \leq 2$, with a sharp fall-off to $\sigma^2_t \sim t^0$ for $V > 2JT$ \cite{PRB2}.

\begin{table}[t]

	\begin{tabular}{|p{1em}|p{8em}|p{16em}|}
	\hline
	\multicolumn{2}{|c|}{\textbf{Type of Noise}} & \textbf{Description}\\
	\hline
A & Timing noise (Aperiodicity) \cite{PRB2} & The position of kick (in time axis) is a uniformly distributed random number drawn from the interval $[T-\delta T,T+\delta T].$ \\
	\hline
B & Amplitude noise \cite{PRA-levy} & Kicks are delivered periodically, but after time intervals randomly chosen from Yule-Simon distribution, the amplitude $V$ is altered by a random amount $\delta V$. \\
 	\hline
C & Timing noise (Missed kicks) \cite{PRL-santh} & The spacing between kicks is a random number drawn from uniform, normal and Poisson distributions and hence some kicks are missed.\\
	\hline
\end{tabular}

\caption{\centering Descriptions of three noisy kicking protocols, labeled A, B and C, used in this work. See Fig. \ref{schematic} for a visual schematic of these protocols.}
\label{tab:kscheme}
\end{table}

\subsection{Classical limit}	
In the classical limit, the Hamiltonian in Eq. \ref{model} exhibits chaotic dynamics since kicking eliminates energy as a conserved quantity. This can be seen in the classical Hamiltonian obtained by taking a continuum limit of the kicked tight-binding model as follows :
\begin{align}
\sum_x \op{x}{x+1} &\rightarrow \exp(-\iota\widehat{p})\nonumber\\
\sum_x \op{x+1}{x} &\rightarrow \exp(\iota\widehat{p}).\\
-J\sum_{i=1}^{L} \left(c_i^\dag c_{i+1} + c_{i+1}^\dag c_i\right) &= -J\sum_x \left(\op{x}{x+1} + \op{x+1}{x}\right),\nonumber\\
  &= -2J \cos(\widehat{p})\\
H_{\text{\footnotesize cl}} = -2J \cos(p)\text{ } + \text{ }V&\sum_{n}\delta(t-T_n)\cos(2\pi\alpha x).
\end{align}
Starting from $H_{\text{cl}}$ we can derive the so-called kicked Harper map \cite{PhysRevLett.67.1377},
	\begin{align}
	p_{n+1} &=p_n + 2\pi\alpha V\sin(2\pi\alpha x_n), \nonumber \\
	x_{n+1} &=x_n + 2J\tau \sin(p_{n+1}).
	\end{align}
Numerical investigations have shown that such a model is chaotic regardless of the value of $\alpha$, leading one to posit that all kicked tight-binding models ought to be quantum chaotic. Numerous studies on this map \cite{PhysRevLett.67.1377,PhysRevLett.68.3826,doi:10.1142/S0217979294000099} have effectively established what is known about the Aubry-Andr\'{e}-Harper model to be true even in the presence of kicking, with the addition of intriguing dynamics such as ballistic transport \emph{and} quantum localization, depending on the location in (semi-classical) phase space \cite{PhysRevLett.67.1377}. These have been corroborated by studies on their tight-binding analogues which have begun to be explored only recently \cite{PRB1,PRB2}.

\begin{figure}[t!]
\includegraphics*[width=2.8in]{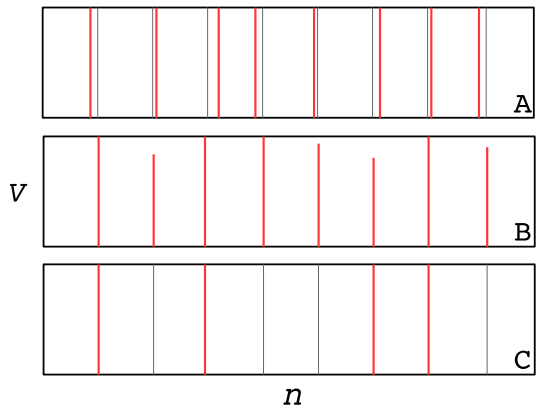}
\caption{Schematic of noisy kicking protocols. Kick amplitudes are shown as a function of integer time $n$. Grey color vertical lines are the positions of kicks had the kick sequence been periodic. Red colour lines indicate the actual kicks imparted for the given type of noise protocol. Labels A, B and C refer to three kicking protocols (see Table \ref{tab:kscheme}).}
\label{schematic}
\end{figure}

\section{Noisy Driving : Protocols and methods}
\label{sec3}
In this section, results related to the dynamical effects that arise from a stochastic perturbation to the driving term in Hamiltonian in Eq. \ref{model} are presented. Indeed, some effects of aperiodicty in the kicking have been studied previously \cite{PRB2}, where it was found that localization persists even in the presence of strong aperiodicity for sufficiently long time. The kicking protocol used was such that the time interval between kicks was drawn uniformly from an interval $(T-\delta T,T+\delta T)$. Against this backdrop, in this paper, the AAH model has been subject to stronger forms of noise protocols in order to probe the range of phenomena that driving, periodic or noisy, can elicit.
In this paper, noise protocols were implemented by modifying the periodic potential in Eq. \ref{model} to
\begin{equation}
\sum_{n} ~ (V + \delta V_n) ~ g_n ~ \delta(t-n \tau) \sum_{i=1}^{L} V_i ~c_i^\dag c_i.
\end{equation}
These noisy protocols are listed in \cref{tab:kscheme}. The case of the periodic kick sequence corresponds to $g_n=1$ and $\delta V_n=0$ for all $n$. Protocol A, shown in Fig. \ref{schematic}(A), corresponds to $g_n=1$ and $\delta V_n > 0$ for all $n$, and its outcome has been reported before \cite{PRB2}. In this protocol, white noise is superimposed on kick times \cite{PRB2} such that, on an average, there is still one kick during every time period $\tau$ though the actual positions of kicks are noisy. In protocol B, visualized in Fig. \ref{schematic}(B), a kick is always delivered at periods of $\tau$ and hence $g_n=1$. However, after time intervals drawn from Yule-Simon distribution given by
\[
f(k) = \beta B(k, \beta+1) \stackrel {k \to \infty}{\Longrightarrow}  k^{-\beta-1},
\]
the amplitude is made noisy ($\delta V_n > 0$) and is drawn from a uniform distribution. In this, $B$ is the Beta
function and $\beta$ is a parameter. In protocol C, $\delta V_n=0$ and $g_n$ takes either 0 or 1. If $g_n=0$ ($g_n=1$), kicks are missed (present). The time interval between successive occurrences of 1 is $K_n \tau$ with $K_n$ being an integer drawn from uniform, Poisson or normal distributions.

\begin{figure}[t!]
\includegraphics*[width=0.45\textwidth]{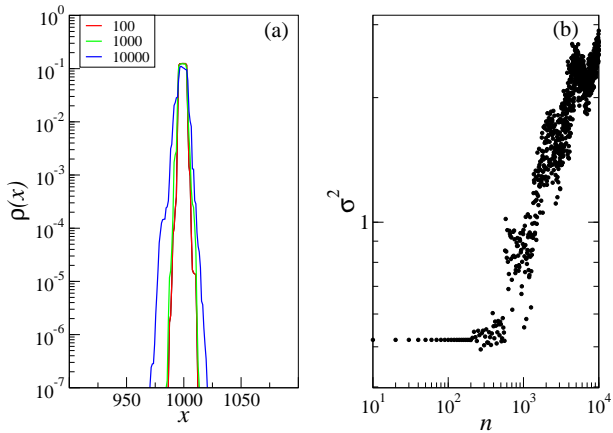}
\caption{Density profile $\rho(x)$ and spread of the wavefunction under protocol B. (a) $\rho(x)$  
at times $n=100, 1000$ and 10000. (b) Width of the time evolving wavefunction (indicated by $\sigma^2$).
In this, timing noise is drawn from a Yule-Simon distribution with 
exponent $\beta=0.75$ and amplitude noise is uniformly distributed in the range $\left[-V/2,V/2\right]$.}
\label{prot1}
\end{figure}
	
In this work, the evolution of a wavepacket, initially localized at the center of the lattice, is characterized by two quantities -- its (noise-averaged) spread $\sigma^2(t)$ and an instantaneous, locally averaged diffusion exponent $\gamma_m$, defined as follows :
		\begin{align}
		\sigma^2(t)&= \sum_{i=1}^{N}(i-N/2)^2\overline{\abs{C_i(t)}^2};\ C_i(t) \equiv \braket{x_i}{\psi(t)}\label{sigdef},\\
		\gamma_m(t) &\equiv \frac{1}{m}\text{ } \sideset{}{'}\sum_{\bar{t}=-\frac{m}{2}}^{\frac{m}{2}}\frac{\log(\sigma^2\left(t+\bar{t}\right)) - \log(\sigma^2(t))}{\log\left(1+\frac{\bar{t}}{t}\right)}\label{gamdef}.
		\end{align}
The utility of $\gamma_m$ lies in its ability to expose local variations in the dynamics. The caveat, then, is that genuine local fluctuations need to be aptly distinguished from noise, especially in the cases where the system is aperiodically driven.
		
The numerical results shown in this paper were implemented using the Armadillo linear algebra library \cite {sanderson2016armadillo,sanderson2018user}. The initial wavefunction, placed at the middle of the lattice of size $L$, 
$\braket{x}{\psi(t=0)} = \delta_{x,L/2}$ is evolved under the action of noisy kicks applied to the Hamiltonian in Eq. \ref{model}. 
If the kicks are periodically applied (noiseless case), then the wavefunction at every time step is acted upon by the unitary operator 
\begin{equation}
	U(\tau^-,0^-) = e^{-\iota\tau\widehat{H}_0} ~ e^{-\iota\widehat{V} \Delta t},
\label{flqop1}
\end{equation}
with $\Delta t =\tau$ only for the strictly periodic case. The superscript \lq{}$-$\rq{} indicates the mapping of the system between two instants immediately before kicks. In noisy cases, the operator $e^{-\iota\tau\widehat{H}_0}$ is applied at every time step while 
the kicking potential $e^{-\iota V}$, $\text{\ }\widehat{V}=V\sum_{j=1}^{L} \cos\left(2\pi\alpha j + \phi\right) c_j^\dag c_{j}$ is applied at times $t_n$ and the time interval $\Delta t$ will be a random variable as per the protocols listed in Table \ref{tab:kscheme}. Apart from these noisy protocols, the case where the kick operator $e^{-\iota\tau \widehat{V}}$ is applied based on the result of a biased coin-toss provides an insight into the role played by the correlations in the kicks. This is discussed in section \ref{sec7}(D). All these are implemented by direct matrix multiplication, since $\widehat{H}_0$ is tridiagonal in the position basis, and can be efficiently diagonalised and hence exponentiated.

\begin{figure}[t!]
\includegraphics*[scale=0.32]{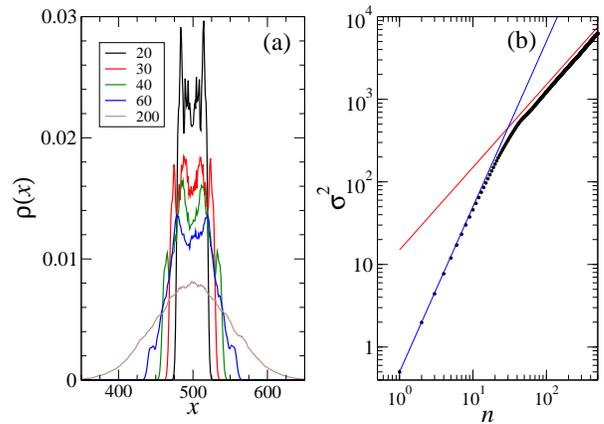}
\caption{Density profile $\rho(x)$  and its evolution under protocol C for parameters $V=1.5$ and $A=50$.
(a) $\rho(x)$  at times $n=20, 30, 40, 60$ and 200. (b) Width of the time evolving wavefunction as a function
of time. The dashed lines have slopes 1 and 2. See text for details of noise imposed on kick sequence.}
\label{prot2}
\end{figure}

\section{Evolution under noisy kicks}
\label{sec4}
\subsection{Noisy kicks versus missing kicks}
For the purposes of the numerical results presented in this section, the parameters 
for the noise-free AAH model are set at $\tau=0.5$ and $J=1$ resulting in 
$V_\text{c} \sim 2J\tau = 1$ at which the delocalization to localization transition takes place. 
To motivate the main results, two broad scenarios arising from the application of 
protocols B and C are displayed in Fig. \ref{prot1} and \ref{prot2}. The density profile of 
the evolving wavepacket and $\sigma^2(n)$ are shown in Fig. \ref{prot1}(a,b) for protocol B. Remarkably, when the model is in the localized phase ($V=1.5$), despite strong amplitude noise with a width equal to $V$, the density profile nearly retains its shape. The corresponding growth of $\sigma^2(n)$ is so strongly muted that it is of $O(1)$ even after $10^4$ time steps. A similar scenario of robust localization emerges if protocol A is applied (not shown here) and this has been reported before in Ref. \cite{PRB2}.
\begin{figure}[t!]
\hspace{-3em}
\includegraphics[width=0.42\textwidth]{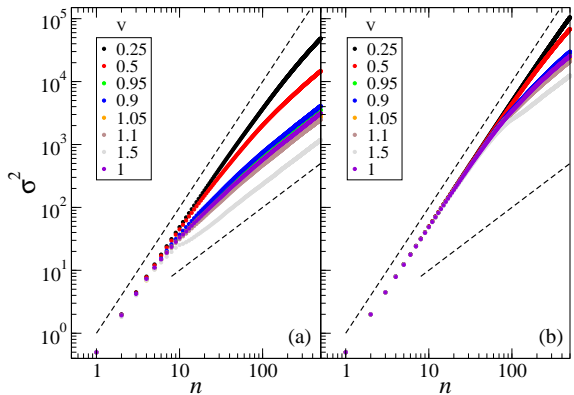}			
\caption{Wavefunction evolution under the effect of missed kicks (protocol C), with the time intervals 
without kicks are drawn from uniform distribution. Width of the evolving 
state $\sigma$ vs. time (symbols) shown for several values of kick strength $V$ with (a) $A=10$ and (b) $A=100$.
For $V>V_c$, a sharp break in the dynamics is observed at $n=n_c \sim A$. The dashed lines have slopes 1 and 2.}
\label{srep}
\end{figure}

Hence, in any protocol that ensures that a kick is consistently delivered at multiples of $\tau$, or 
at least in a small time window around it \cite{PRB2}, the wavepacket spread is weakly subdiffusive, 
indicating the robustness of the localized phase to noise. Even if the amplitude is altered at a random time, 
localization persists for an extensive number of kicks, as evident in \cref{prot1}(a,b). This picture changes 
when a kick is not guaranteed at every timestep as in the case of protocol C displayed in Fig. \cref{schematic}(c).
If $T_N$ is the time at which the $N$-th kick is delivered, then $K_N \equiv \left(T_{N+1}-T_N\right)/\tau$, 
where $K_N>0$ is an integer random variable drawn from a discrete distribution $F(K)$. In the rest of the
paper, we shall call $F(K)$ the waiting time distribution since it represents the time interval between
successive appearance of kicks. 

As we show in the next section, the character of diffusion undergoes a sharp change at a timescale 
decided by the specifics of the noise process. This is observed across different types of noise 
superimposed on the standard kick sequence. Figure \ref{prot2}(a) shows the evolution of density profile
due to noisy protocol C. A global picture is seen in Fig. \ref{prot2}(b).
At short timescales, the dynamics can display {\bf subdiffusive} to ballistic growths, and tends asymptotically 
towards diffusive dynamics. This crossover from growth with slope 2 (in log-log plot) to slope 1 is 
observed in Fig. \ref{prot2}(b). If $n_c$ represents this 
crossover timescale, then superdiffusive growth takes place for times $n < n_c$, and becomes 
diffusive for $n > n_c$. The numerical value of $n_c$ is dependent on the properties of superimposed noise.
Given that localization is robust even if noisy kicks are imparted once every kicking period,
in the subsequent sections, we focus on the wavefunction evolution under the effect missed kicks (protocol C).
In the limit of $t \to \infty$, the unbounded diffusive growth continues though in practice it is interrupted by
the finite size of the lattice.
		
\section{Dynamics under missing kicks}
\label{sec5}
	In this section, the diffusive effects induced by the missing kicks, protocol C, are studied. 
To begin with, since $K \in \mathbb{Z}$, we consider the case of $F(K)$ being a discrete uniform
distribution $U(K;A)$, with $K \in \left[1,A\right]$ defining the support for the distribution.
In this case, a ballistic growth at initial times and an asymptotic diffusive growth is generically observed.
The simulations that result from the Hamiltonian in Eq. \ref{model}, shown in Fig. \ref{srep}, reveal the 
existence of a crossover timescale $n_c$. In Fig. \ref{srep}(a), $\sigma^2$ is shown as a function of $n$
for several values of kick strength $V$ for $A=10$. At short time scales of $n < n_c$, the wavefunction spread is 
ballistic as the increase in $\sigma^2$ is parallel to dashed line with slope 2. For $n > n_c$, wavefunction
spread becomes diffusive, as evident in $\sigma^2(n)$ becoming parallel to a line with slope 1 (Fig. \ref{srep}(a)).
This scenario of transition from ballistic growth at short times to asymptotic diffusive growth repeats 
for $A=100$ as well, as shown in Fig. \ref{srep}(b).

\begin{figure}[t!]
	\hspace{-2em}
	\includegraphics*[width=0.42\textwidth]{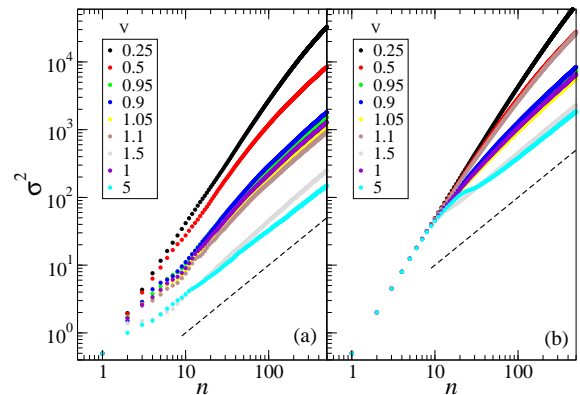}
	\caption{Wavefunction evolution under the effect of missed kicks (protocol C), with waiting times
drawn from Poisson distribution. Width of the evolving state $\sigma$ vs. time (symbols) shown for several 
values of $V$ for (a) $\lambda=0.5$ and (b) $\lambda=10$. Note the sharp transition point for $\lambda=10$. 
Dashed lines have slopes 1 and 2.}
	\label{srep2}
	\end{figure}

When the kicking is in the delocalized regime, $V < V_c$, then the transition is smooth and is corroborated in
Fig. \ref{srep}(a). However, in the cases when $V \gtrsim V_c$, the transition suddenly sharpens, and one can notice distinct regimes in the dynamics. In this case, a distinct crossover point between the two regimes is observed. A common feature is that regardless of the number of distinct regimes, transitions between them are always sharp and hence we can label them as being distinct. In particular, for $V > V_c$ , the transition to asymptotic diffusive behaviour always 
occurs at $n = n_c \approx A$. This is borne out by the numerical results in Fig. \ref{srep}(a,b).
The asymptotic diffusive behaviour is expected as a generic feature of decoherence \cite{Zurek2007}, 
which takes place when noise is introduced into a system \cite{PhysRevLett.53.2187,PhysRevLett.67.1945}.

\begin{figure}[t!]
	\hspace{-2.5em}
	\includegraphics*[width=0.42\textwidth]{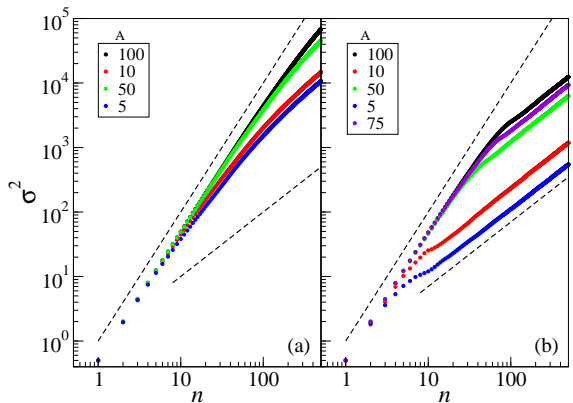}
	\caption{Growth of density profile in the delocalized and localized regime as
the parameter $A$ of the waiting time distribution $U(K;A)$ varies. In this, (a) $V=0.5$ and, (b) $V=1.5$ are fixed, 
but $A$ is varied. Abrupt transitions from superdiffusive to diffusive growth is seen only in the localized
regime with $V>V_c$. Dashed lines have slopes 1 and 2.}
	\label{Vfix_Avar}
	\end{figure}

If the waiting times are Poisson distributed, {\it i.e.},  $F(K)=P(K;\lambda)$, 
where $\lambda$ is the mean of the distribution, a scenario similar to that of uniform
distribution emerges. For $\lambda<1$, which physically implies that kicks are rarely missed, the dynamics 
of the almost periodically kicked Hamiltonian gives rise to approximately ballistic initial growth until 
asymptotic diffusive growth sets in at later times. The timescale at which this sets in is much 
longer when $V<1$ than if $V>1$. These features are observed in the simulation results in Fig. \ref{srep2}(a) 
for which $\lambda=0.5$. In all the cases, asymptotic diffusive growth sets in as $n >>1$.
For large $\lambda$, as shown in Fig. \ref{srep2}(b), there is an initial ballistic growth and it
crosses over to normal diffusion, as in the case of uniform noise. The transition point $n_c$ 
from superdiffusive to diffusive growth is sharp if $V>1$, and it is smooth if $V<1$.

\begin{figure}[t!]
		\hspace{-2.5em}
		\includegraphics*[width=0.42\textwidth]{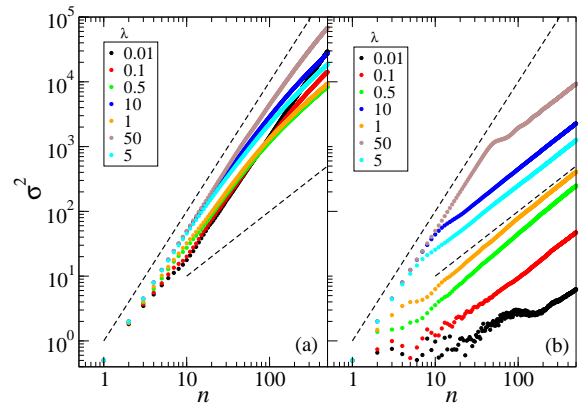}
		\caption{Growth of density profile in the delocalized and localized regime with the change in
the parameter $\lambda$ of the waiting time distribution $P(K;\lambda)$. In this, (a) $V=0.5$ and, (b) $V=1.5$ are fixed, 
but $\lambda$ is varied. Sharp transitions from superdiffusive to diffusive growth is seen only in the localized
regime with $V>V_c$ and $\lambda >> 1$. Dashed lines have slopes 1 and 2.}
		\label{Vfix_Lvar}
		\end{figure}

What is the nature of transition if the parameter characterising the waiting time distribution is varied ?
The qualitative picture portrayed in Fig \ref{srep} does not change even if the parameter $A$ in the uniform waiting time distribution $U(K;A)$ is varied. As displayed in Fig. \ref{Vfix_Avar}, varying $A$ does not modify the initial superdiffusive regime and the asymptotic diffusive regime. Its effect is to simply shift the time at which the sharp transition is observed for $V>V_c$ (Fig. \ref{Vfix_Avar}(a)), and to change the timescale on which asymptotic 
diffusion sets in for $V<V_c$ (Fig. \ref{Vfix_Avar}(b)).
The distinct dynamical behaviour upon variation of $\lambda$ is evident in Fig. \ref{Vfix_Lvar}. No sharp
transitions are observed for $V<1$, irrespective of the value of $\lambda$. In this case, all the transitions
are smooth as seen in Fig. \ref{Vfix_Lvar}(a). 
For large $\lambda$, longer mean waiting times ensure that the ballistic regime stays on for longer timescales.
On the other hand, for $V>1$ sharp ballistic to diffusive transitions are observed if $\lambda > 1$ 
(Fig. \ref{Vfix_Lvar}(b)). In section \ref{sec7}, we provide a justification for these observations
and also for the qualitative similarity of the dynamics with Poisson noise $\left(\lambda\gtrsim2\right)$ to 
that of uniform noise. Thus, the parameters $A$ and $\lambda$ effectively tune the separation of timescales
with distinct growth regimes. Further, similar results as in Figs. \ref{srep}-\ref{Vfix_Lvar} were also 
observed (not shown here) when $K_n$ were chosen as the integer parts of a normally distributed variable.

\section{Translationally Invariant Models}
\label{sec6}
In the light of the results presented above, one significant question of interest
would be the dependence of the observed phenomena on the lack of translational invariance in the model. To address this, the parameter 
$\alpha$ in the onsite potential $V_i$ in Eq. \ref{model} is tuned from rational to an irrational number. 
A standard way to do this is to consider a sequence $\alpha_i$ such that $\alpha_i = F_i/F_{i+1}$, 
where $F_i$ is the $i^\text{th}$ Fibonacci number. Then, $\lim\limits_{i\rightarrow\infty}\alpha_i = \frac{\sqrt{5}-1}{2}$. 
When $\alpha = \frac{p}{q}$, where $p$ and $q$ are integers, the Hamiltonian in Eq. \ref{model} becomes 
translationally invariant and will have $q$-bands under periodic driving. In this section, we simulate 
the wavefunction evolution for the $q$-band model and obtain $\sigma^2$ as a function of time $n$.

	\begin{figure}[t!]
	\hspace{-2.5em}
	\includegraphics*[width=0.5\textwidth]{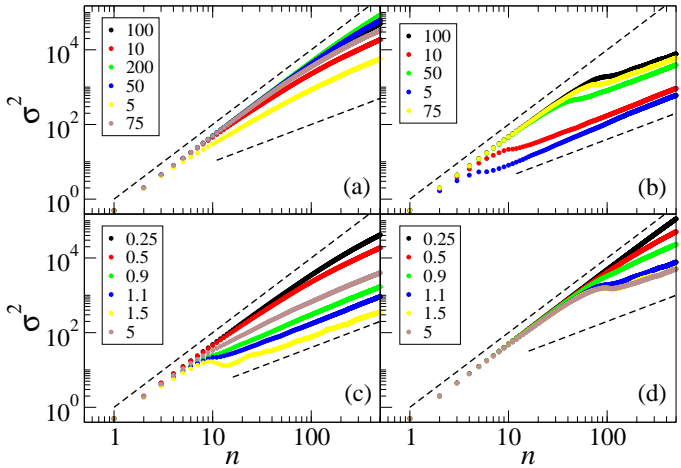}
	\caption{Diffusion in 2-band model with uniformly distributed waiting times between kicks.
	The parameters are (a) $V=0.5$ and (b) $V=1.1$ for various values of $A$, and 
	(c) $A=10$ and (d) $A=100$, for various values of kicking strength $V$.  Dashed lines have slopes 1 and 2.}
	\label{2B_sig_uni}
	\end{figure}

\begin{figure}[t!]
		\hspace{-2.5em}			
		\includegraphics[width=0.5\textwidth]{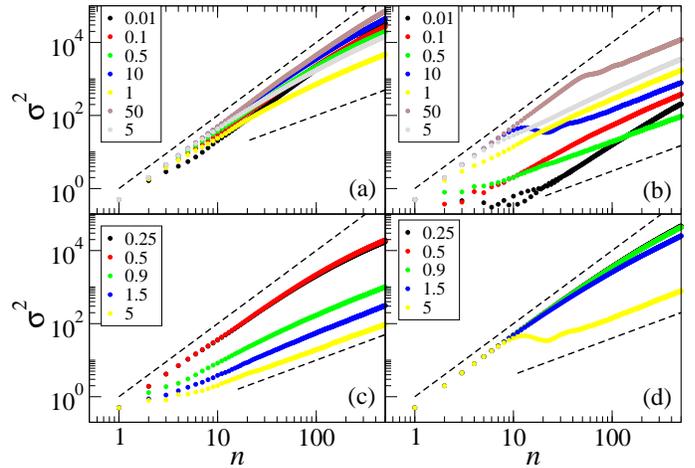}
		\caption{Diffusion in 2-band model with Poisson distributed waiting times between kicks.
	The parameters are (a) $V=0.5$ and (b) $V=1.1$, for various $\lambda$, and (c) $\lambda=0.5$ and (d) $\lambda=10$, for various $V$. Dashed lines have slopes 1 and 2.}
		\label{2B_sig_poi}
		\end{figure}

Firstly, as Figs. \ref{2B_sig_uni} and \ref{2B_sig_poi} show, a transition from ballistic to diffusive growth of $\sigma$ persists even in the translationally invariant models. Further, the transition is sharp under identical parametric regimes, {\it i.e.,} $V>V_c$, considered earlier in Figs. \ref{Vfix_Avar} and \ref{Vfix_Lvar}. As the system is switched from a free model to a $q$-band model, where $q$ is a small integer, it is surprising to observe a sharp break connecting these two regimes. In the simulations with 2-band model and noisy protocol C with uniform waiting time distributions, sharp transition is absent if $V<V_c$ whereas it is clearly visible if $V>V_c$ as seen in Figs. \ref{2B_sig_uni}(a,b,c,d). In the case of Poisson distributed waiting times shown in Fig. \ref{2B_sig_poi} a sharp transition is seen if $\lambda >> 1$. The asymptotic diffusive dynamics in all the cases arises due to decoherence. Such dynamics is generically seen across all the $q$-band structures we have simulated for $2\leq q\leq10$. Hence, the results shown in Figs. 4 to 7 are also valid for models with translational invariance.

In general, what determines if the nature of transition will be sharp or smooth? Further, in certain windows of the strength of kicks $V$, a drastic slowing down in the growth of $\sigma$ at the transition point is observed. This is accompanied by the diffusion exponent almost reaching zero and, in certain cases, even falling further. To explain the observed features in Figs. \ref{srep} to \ref{2B_sig_poi}, we further study the band structure of the $2$-band model in greater detail.
	
		\subsection{Band Structures : kicking induced flat bands}
A first striking observation from the band structures is that, for the model in Eq. \ref{model-intpart}, 
kicking alone can be used to engineer flat-bands and hence, the localization of excitations. Lattice models
that exhibit flat bands have been of considerable interest in recent times \cite{PhysRevB.82.104209,annphys.529.1600182, PhysRevB.99.235118}. An effective 
Hamiltonian which gives rise to the same dynamics as that of the periodically driven model over one time 
period can be obtained from the Floquet operator using the Baker-Campbell-Hausdorff expansion. In a na\"{\i}ve 
weak-potential approximation at high kick frequencies, one simply obtains an effective Hamiltonian
$H_{\mathrm{eff}} = \widehat{H}_0 + \widehat{V}/\tau$, leading to a tight-binding model with the 
potential rescaled by $\tau$. However, for no on-site potential of finite strength can there be flat bands 
in static $q$-band tight-binding models. Analytically, this can be seen for the simple cases of the 2, 3 and 4 
band models. Concretely, the dispersion relation for a 2-band model with onsite potentials $\pm {\mathcal U}$ 
is given by \cite{markos2008wave}
		\begin{equation}
	E(k) = \pm\sqrt{\mathcal{U}^2 + 4J^2 \cos^2\left(\frac{ka}{2}\right)}
		\label{disp2},
		\end{equation}
where $a$ is the lattice spacing, and $J$ is the hopping amplitude. Only in the limit where the hopping 
is completely suppressed and there are on-site potential terms alone, is the spectrum flat.

\begin{figure}
		\includegraphics*[width=0.28\textwidth,angle=-90]{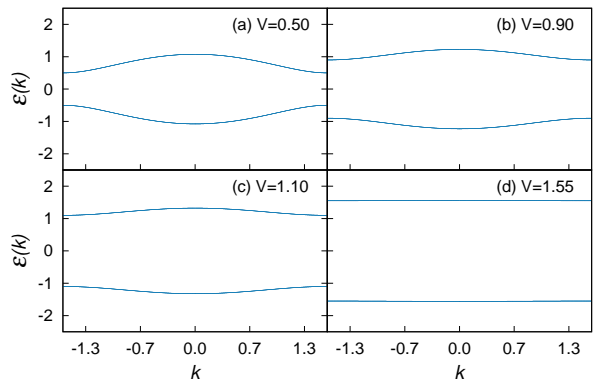}
		\caption{Band structures obtained from simulations for the 2-band model. Note that the flat bands 
are obtained for $V=\pi/2 \approx 1.55$.}
		\label{bandstruct-2B}
		\end{figure}

Evidently, this is not the case for the kicked AAH model in Eq. \ref{model}.  Figure \ref{bandstruct-2B} shows the band structure
for the 2-band model. As seen in Fig. \ref{bandstruct-2B}(d), flat bands can indeed be realised with finite on-site potential in 
the presence of kicking. By diagonalising the Floquet operator, the quasi-energies $\epsilon(k)$ can be expressed as
	\begin{align}
	\epsilon(k) = \pm\cos^{-1}\left\{\cos\left(2J\tau\cos k\right)\cos V \right\}
        \label{dispf2}.
	\end{align}
Thus, flat-bands exist for $V=\left(2l+1\right)\frac{\pi}{2}, \text{\ } l\in\mathbb{Z}$ and the 
kicking has shown a relatively straightforward mechanism to construct flat bands. This is confirmed
in Fig. \ref{bandstruct-2B}(d) where flat bands appear
for $V=\pi/2 \approx 1.55$. At such parameter values, where the allowed bands are exactly flat,
an abrtupt transition point is indeed observed (Fig. \ref{2B_sig_uni}(c,d)).
		
However, as simulation results reveal, {\sl perfectly} flat bands 
are not required in order to observe either a sharp break or a slowing down in the dynamics. For instance,
sharp transitions begin to appear approximately for $V>V_c$ as seen in Fig. \ref{2B_sig_uni}(c,d).
Hence, in the next section, we provide analytical support to the observations in sections \ref{sec3} to \ref{sec6}
using 2-band model. In particular, we also analyse the question, how \lq\lq{}flat\rq\rq{} is \lq\lq{}flat enough\rq\rq{} ? 
Moreover, are there other transitions that one can expect to observe that accompany 
the appearance of a break in the dynamics ?

\section{Analysis}
\label{sec7}
In this section, analytical justification is presented, mainly, for the case 2-band model.
Firstly, $\gamma_4$, as defined in \cref{gamdef}, is computed to obtain $n_c$, 
the time at which sharp transition takes place in the dynamics. Secondly, as a further evidence for
sharp transtion at time $n=n_c$, a na\"{\i}ve scaling of $\log(\sigma^2)$ is presented in the 
case of uniform noise. Finally, the role of noise correlations in the sharpness of 
the transition is elucidated.

	\subsection{Calculation of $\gamma_4$}
The time $n_c$ at which the dynamics sharply transitions from superdiffusive to 
diffusive is obtained by calculating the averaged local diffusion exponent $\gamma_m$, and 
observing the time at which sharpest drop occurs. As is evident from Figs. \ref{Vfix_Avar}-\ref{Vfix_Lvar}, 
the transition from initially superdiffusive to diffusive growth is smooth in the 
delocalized regime of kick strength $V<V_c$. This transition becomes sharp as one approaches and 
crosses into the localized regime. In our investigations, there does not appear to be any qualitatively 
different behaviour at the critical point $V=V_c$.
		
\begin{figure}[t]
		\hspace{-2em}
		\includegraphics*[width=0.43\textwidth]{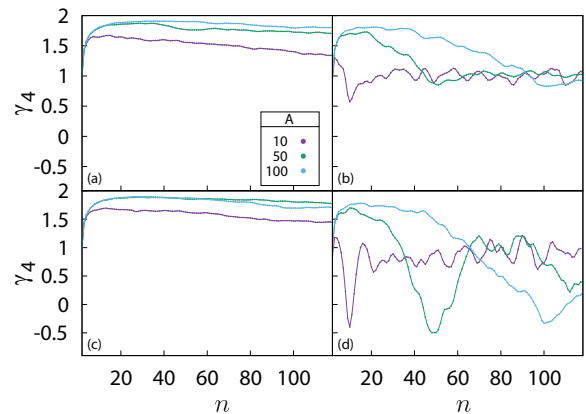}
		\caption{$\gamma_4$ vs. $n$ for noisy models with waiting times drawn from a uniform distribution $U(K)$, 
                with $K \in [1,A]$. (a,b) AAH model, and (c,d) 2-band model. The parameters are (a,c) $V=0.5$, (b,d) $V=1.5$.}
	\label{g4-unidist}
	\end{figure}

		In \cref{g4-unidist}, for the case of AAH model (\cref{g4-unidist}(a,b)) and 
2-band model (\cref{g4-unidist}(c,d)), $\gamma_4(n)$ is shown for uniformly distributed waiting times 
$U(K)$ with $K \in \left[1,A\right]$. Clearly, a sharp dip in 
$\gamma_4$ occurs exactly at $n=A$ when $V>V_c$. In contrast, for $V<V_c$, a smooth transition occurs from 
the superdiffusive and diffusive regime.

In fact, the dips in $\gamma_4$ in general seem to be far more steep in the translationally invariant 
case than in the AAH model, across different values of $V$. As expected, the flatter the bands, 
the sharper the dips in $\gamma_4$. This provides a quantitative evidence to associate flat bands with
sharp transition in diffusive regimes.
	
	\subsection{Scaling of $\sigma ^2(t)$}
		The scaling arguments are presented in this section to corroborate the fact that the 
time of transition is $n_c \sim A$, when $K$ is drawn from a uniform distribution $\left[1,A\right]$. 

\begin{figure}[t]
		\hspace{-2em}
		\includegraphics*[width=0.5\textwidth]{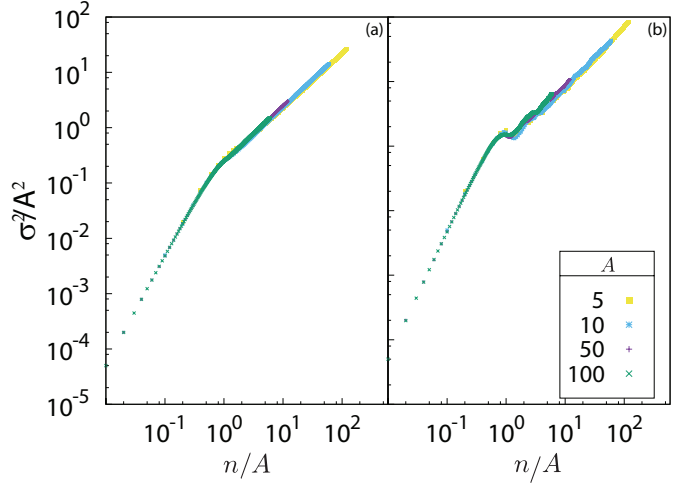}
		\caption{Scaling of $\sigma^2$ with the $A$ for the case of uniformly distributed waiting times with $V=1.5$. 
		(a) AAH model, and (b) 2-band model. Note the excellent data collapse after scaling using Eq. \ref{scalingeqn}.}
		\label{scaling}
		\end{figure}

		It can be posited that $\sigma^2(n;A)$ can be expressed in terms of a function $\sigma^2_0$, 
		defined as
		\begin{align}
			\sigma_0^2(\bar{n})=\begin{cases}
			\bar{n}^2 & 0<\bar{n}<1 \\
			\bar{n} & \bar{n} > 1
			\end{cases},\\
			\sigma^2(n;A) = A^2\sigma_0^2\left(\frac{n}{A}\right).
 			\label{scalingeqn}
		\end{align}
In order to demonstrate this scaling, (the logarithm of) the function 
$f_A(\bar{n})\equiv \frac{1}{A^2}\sigma^2(A\bar{n})$ is plotted for different values of $A$ in 
\cref{scaling}. It can be seen from this figure that $f_A(\bar{t})\sim\sigma_0^2(\bar{t})$ for all $A$.
Taken together, for uniformly distributed waiting times, Figs. \ref{bandstruct-2B}-\ref{scaling} display the 
central result that at parametric regimes at which flat bands appear,
the transition from superdiffusive to diffusive growth of evolving states is sharp, and the transition
point follows a excellent scaling with respect to a parameter $A$ that controls the strength of noise.

	\subsection{Mechanisms for time of transition and diffusion}
In this section, analytical support is provided for associating flat bands with the sharp superdiffusive to diffusive transition in the case of 2-band model. To begin, the following
three probabilities are defined:

\begin{tabular}{p{0.06\textwidth}p{0.4\textwidth}}
	$\mathcal{P}(n)\equiv$& the probability of a kick occurring at the $n^{th}$ time step\\
	$P_N(n)\equiv$&\hspace{7pt}the probability of the $N^{th}$ kick occurring at the $n^{th}$ time step\\
	$Q(n;N)\equiv$&\hspace{17pt}the probability that exactly $N$ kicks have been received before the $n^{th}$ time step\\
\end{tabular}
These quantities are related to each other as
\begin{align}
\mathcal{P}(n) &= \sum_{N=1}^nP_N(n)\\
Q(n;N) &= \sum_{n'=N}^n\left(P_{N}(n') - P_{N+1}(n')\right).
\end{align}
If $t_i$, $i=1,2, \dots N$, denote the time intervals between successive kicks on the time axis,
then the calculation of $P_N(n)$ amounts to finding the number of ways one can choose 
$\lbrace t_1, t_2, \ldots, t_N\rbrace$ such that $\sum\limits_{i=1}^{N} t_i = n$. Thus
\begin{equation}
\label{PNdef}
P_N(n) = P_N\left(\sum\limits_{i=1}^{N} t_i = n\right)
\end{equation}
 For the specific case of $F(K) = U(K;A)$, note that $P_N(n) = 0$ if $N<\left[\frac{n}{A}\right] + 1$, and by construction, $P_N(n) = 0$ if $N>n$ for all cases of $F(K)$. The normalisation condition is
 \begin{equation}
\label{norml}
\sum_{n=N}^{NA}P_N(n) = 1
 \end{equation}
from which one can see that
 \begin{align}
 Q(n\geq (N+1)A;N) &= \sum_{n'=N}^n\left(P_{N}(n') - P_{N+1}(n')\right)\nonumber\\
 &=\sum_{n'=N}^{NA}P_{N}(n') - \sum_{n'=N+1}^{(N+1)A}P_{N+1}(n')\nonumber\\
 &=1 - 1 = 0.
 \end{align}
The last form uses the normalization in Eq. \ref{norml}. In all cases, $\mathcal{P}(n)$ does depend 
on $n$, at least for small $n$.

This discussion begins by considering what happens to an initial state $\psi_0$ under the
dynamics of the 2-band model after the system receives exactly one kick. 
The quantity of interest, then, is
	\begin{align}
		\widetilde{\sigma}^2(t_1,t_2,V) &\equiv \expval{e^{\iota t_1 \widehat{H}_0}e^{\iota \widehat{V}}e^{\iota t_2 \widehat{H}_0}\hat{x}^2 e^{-\iota t_2 \widehat{H}_0}e^{-\iota \widehat{V}}e^{-\iota t_1 \widehat{H}_0}}{\psi_0};\nonumber\\
	\ket{\psi_0}&\equiv\frac{1}{\sqrt{L}}\sum_{k}\ket{k}.\label{sigtil}
	\end{align}
where $|k\rangle$ are the momentum states. In the momentum basis, the evolution operator decomposes 
into $2\cross2$ subspaces, in which,  $\widehat{H}_0$ is diagonal and 
$x^2\rightarrow\sum_{k_1,k_2}\partial_{k_1}\partial_{k_2}\delta(k_1-k_2)$.	
The Floquet matrix in a single $2\cross2$ $k$-subspace is
		\begin{align}
		F_k(V,T) = \begin{pmatrix}
		e^{\iota 2JT\cos k}\cos V &-\iota e^{\iota 2JT\cos k}\sin V\\-\iota e^{-\iota 2JT\cos k}\sin V&e^{-\iota 2JT\cos k}\cos V
		\end{pmatrix}.\label{flqop}
		\end{align}
After some calculations whose details are given in Appendix \ref{appsec1}, we obtain
	\begin{equation}
		\widetilde{\sigma}^2(t_1,t_2,V) = 2\left(J\tau\right)^2\left(t_1^2+t_2^2 + 2 \cos(2V)t_1t_2\right).
	\end{equation}
For perfectly flat bands $V=\frac{\pi}{2}$, and this reduces to
	\begin{equation}
	\widetilde{\sigma}^2(t_1,t_2,\pi/2) = 2\left(J\tau\right)^2\left(t_1-t_2\right)^2.
	\end{equation}
This can be interpreted as follows. An impulse with a flat band Hamiltonian leads to a 
reversal of the velocity of the wavepacket. Concretely, consider the case where $F(K)=U(K;A)$. 
The probability that the system has received only one kick by the time $n=A$ steps have 
elapsed is
 \begin{equation}
 Q(A;1) = \left(\frac{1}{2} + \frac{1}{2A}\right) > \frac{1}{2}.
 \end{equation}
Under the assumption that the system is faced with this, the most probable scenario for $\sigma$ would be
\begin{equation}
\sigma^2(A\tau) \sim \frac{\left(J\tau\right)^2}{3}A(A+2) \sim \frac{1}{6}\widetilde{\sigma}^2(A,0,0).
\end{equation}
The leading order term here shows that while the coefficient has reduced, the dynamics is still ballistic. However, when one considers time steps $n>A$, it is most likely that the system will have seen at least 2 kicks (details of
the calculation in Appendix \ref{appsec2}) and its probability can be expressed as
\begin{equation}
Q(n>A;N\geq 2) = 1 - Q(n>A;1) - Q(n>A;0) > \frac{1}{2}.
\label{prob2g}
\end{equation}
\begin{figure}[t]
	\hspace{-2em}
	\includegraphics*[width=0.5\textwidth]{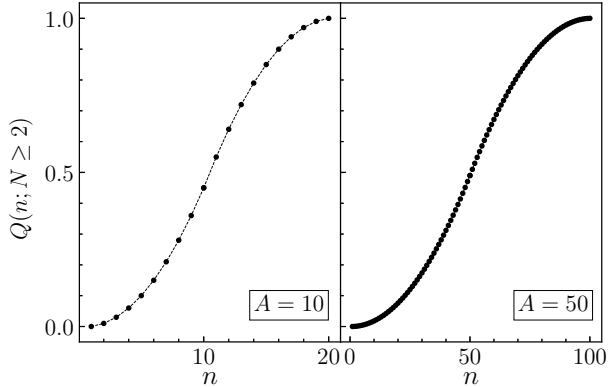}
	\caption{Cumulative mass function $Q(n;N\ge 2)$ for $F(K) = U(K;A)$. The transition happens at 
         time $n_c$ such that $Q(n;N\ge 2)=\frac{1}{2}$. This figure shows that $n_c \sim A$.}
	\label{Qunif}
\end{figure}
As more kicks are imparted, the particle is subject to an increasing number of velocity reversals, 
and it executes a random walk, leading to diffusive behaviour.

Based on this argument, we posit that 
the break-time $n_c$ is the time at which $Q(n;N\geq2)$ -- a monotonic, increasing function -- exceeds $\frac{1}{2}$. Then, for $n > n_c$, the system has most likely received more than one kick. The quantity of interest $Q(n;N\geq 2)$ is shown to be the Cumulative Mass Function (CMF) for the distribution $P_2(n)$ in Appendix \ref{appsec2}. Thus, the posited break-time is the median of the distribution $P_2(n)$ which is $A$, for the case of uniform noise and $\sim\lambda$ for $\lambda>>1$, in the Poisson case.

The CMF for the case of uniformly distributed waiting times is shown in Fig. \ref{Qunif}, for two different values of $A$. Using its expression (obtained in Eq. \ref{QappB} in Appendix \ref{appsec2}), we obtain $n_c = A$, the value at which CMF becomes larger than $\frac{1}{2}$.This estimate for $n_c$ agrees with the time at which sharp transition is observed in Fig. \ref{Vfix_Avar}(b).

\begin{figure}[t!]
	\hspace{-2em}
	\includegraphics*[width=0.5\textwidth]{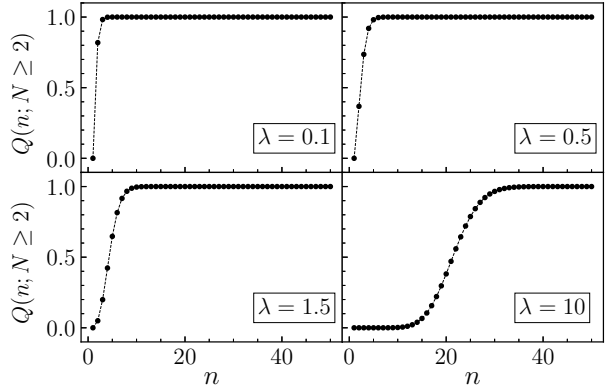}
	\caption{Cumulative mass function $Q(n;N\ge2)$ for $F(K) = P(K;\lambda)$. Notice that $n_c$ is barely greater than 1 for $\lambda\leq1$, explaining why the break is quite abrupt for $\lambda>1.5$}
	\label{Qpois}
\end{figure}

Similarly, for the case of Poisson waiting time distribution, the CMF is shown in Fig. \ref{Qpois} for various values of mean waiting time $\lambda$. In this case, the median of $P_2(n)$ approximately corresponds to the time at which transition to diffusive dynamics takes place. This can be clearly seen for the cases of $\lambda=5,10$ upon comparing the median with the actual transition observed in
Fig. \ref{Vfix_Lvar}(b). Further, the median occurs at $n>1$ only when $\lambda\gtrapprox0.7$, and 
at $n \geq 3$ only when $\lambda\gtrapprox1.25$, which explains why the break is not observable 
in Fig. \ref{Vfix_Lvar}(b) for $\lambda \leq 1$.

At large times $n>>1$, regardless of the form of the $F(K)$, the contribution to $\mathcal{P}(n)$ comes from $P_N(n)$ where $N$ is large. By the central limit theorem and Eq. \ref{PNdef}, $P_N(n)$ tends to a normal distribution that depends only on the properties of $F(K)$, and thus, $P(n)$ is no longer dependent on $n$. This implies that the velocity of the particle is reversed with a fixed probability $P_{\infty} = P(n\to\infty)$ at each time step. Since the wavepacket evolves with that velocity 
for one time step, this is akin to a 1-dimensional classical discrete-time random walk. In this case, it is 
known that $\expval{x^2(t)}\sim t$, and hence a linear increase in the square of the width of the wavepacket
for $n >> 1$ can be expected.

	\subsection{Role of $n$ dependence of $\mathcal{P}(n)$}	
\begin{figure}[b]
	\hspace{-2em}
	\includegraphics[width=0.5\textwidth]{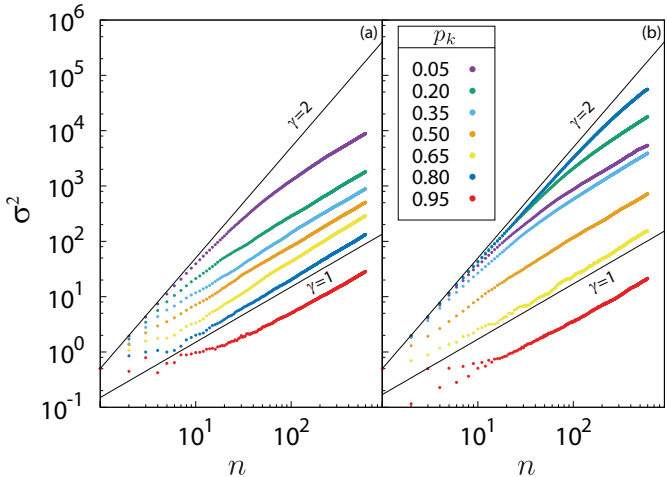}
	\caption{$\sigma^2$ vs.\ $t/\tau$ at $V=1.57$ for coin-toss noise in the (a) AAH model and (b) 2-band model}
	\label{stelv1.57MGR}
\end{figure}

	From the arguments made in the previous sub-section, it is seems that the time dependence of $\mathcal{P}(n)$ at short times plays a crucial role in the appearance of a sharp transition time $n_c$. In order to test this, the AAH model in Eq. \ref{model} is evolved and kicks are applied  depending on the outcome of a biased coin toss, with various (time-independent) biases. $\mathcal{P}(n)$ is now constrained to be independent of $n$, as a result, and is denoted by $p_k$. In the 2-band case, the quantity $\sigma^2$ has an exact solution in terms of a $4\cross4$ \lq\lq{}disorder matrix\rq\rq{}, as shown in Appendix \ref{calc3}. The final result turns out to be
	\begin{align}
	\sigma^2(N)=\frac{1}{L}\sum_{i,j}\left[A(N)+B(N)+C(N)\right]_{i,j}.
	\end{align}
 	where $A(N), B(N)$ and $C(N)$ are defined in the Appendix \ref{calc3}. Note that the indices in brackets are not the true indices of the $4\times4$ matrix $D$. The recipe to switch between the two type of indices is by employing the Kronecker product and is given in Appendix \ref{calc3}. The numerically simulated results for the coin-toss model applied to the Hamiltonian in Eq. \ref{model} and for the 2-band model are, respectively, in Fig \ref{stelv1.57MGR}(a) and \ref{stelv1.57MGR}(b). Notice the absence of sharp transitions even for $V=1.57$ (at which sharp transitions were observed in the case of uniform and Poisson distributed waiting times). This illustrates another facet of wavefunction spreading -- the smooth nature of the crossover from the superdiffusive to the diffusive regime, even at a value of $V$ that produces sharp break-points for other kick sequences with time-dependent $\mathcal{P}(n)$ seems to suggest that this time dependence is crucial to the sharp change observed in the dynamics. This time dependence manifests in form of correlation for the times at which kicks are imparted, in that $\mathcal{P}(n|n') \neq \mathcal{P}(n)\mathcal{P}(n')$. Indeed, there has been work on continuously and stochastically perturbed models where the  has played a non-trivial role in engineering such a transition \cite{PhysRevLett.119.046601}.

\section{Conclusions}
\label{sec8}
This work had focussed on how an initially localized wavepacket evolves under the action
of noisy kick sequences applied to a lattice model with quasi-periodic onsite potential, 
namely, the Aubry-Andr\'{e} model. This system is well known to display localization to
delocalization transition even in the one-dimensional case, in contrast to the
Anderson model. Periodically driving such systems is known to either preserve
localization or induce delocalization depending on the choice of parameters.
This work presents an extensive study of effects due to noisy kicks and we place it
in the context of the current interest in a variety of Floquet engineering schemes
in quantum systems.

In summary, three distinct noisy kick sequences are considered as shown in Fig. \ref{schematic}. The first two sequences deliver one kick on an average in every period, in which case the localized phase appears to be sufficiently robust to application of noise. However, if the kick sequence is chosen such as to miss certain kicks entirely, then the results are quite dramatic. Most of this paper discusses the effects due to randomly missed kicks. Under these conditions, it is found that the dynamics of an initially localized wavepacket can be made ballistic for arbitrarily long times by tuning a parameter characterising the noise process. Following the ballistic regime, there is a sharp transition to diffusive behaviour. In particular, the transition is not always smooth. This was observed both in translationally invariant as well as disordered models; the former required some degree of flatness to the bands. 
We argue and show, using a 2-band model, that the emergence of diffusive behaviour in the long time limit is intrinsically linked to velocity reversals which leads to classical random walk type behaviour. More importantly, by using an uncorrelated kick sequence for comparison, it is shown that correlations in the noise play a crucial role in facilitating the sharpness of this transition, which is determined by the parameters characterising the noise realizations.

It would be interesting to obtain a more rigorous analytic backing for the criterion to find the ballistic to diffusive transition timescale $n_c$. Moreover, the observation of these phenomena in the presence of interactions is yet to be explored in this model, but results in countinuously (stochastically) driven models that could serve as guides to future studies \cite{PhysRevLett.119.046601}.

\acknowledgments
We are grateful to G. J. Sreejith for helpful discussions and comments on the thesis that led to this work. One of the authors (MSS) would like to acknowledge the financial support from Science and Engineering Research Board, Govt. of India, through MATRICS grant MTR/2019/001111.
	
	\appendix
	\section{Calculation of $\mathbf{\widetilde{\sigma}^2\left(t_1,t_2,V\right)}$}
	\label{appsec1}
	
	Let $U_{k1} \equiv F_k(V,t_2)F_k(0,t_1)$, where $F_k(V,T)$ is the same as in \cref{flqop}. $\widetilde{\sigma}^2\left(t_1,t_2,V\right)$, as defined in \cref{sigtil}, can equivalently be written as
	\begin{align}
	&\frac{1}{L}
	\sum_{\substack{\{i_n,j_n\} \\
			{k_1,k_2}}} \mel{i_0}{U^\dagger_{k_2}}{i_1}\mel{j_1}{U_{k_1}}{j_0} \pdv{\delta\left(k_1-k_2\right)}{k_1}{k_2}\nonumber\\
=&\frac{1}{L}
	\sum_{\substack{\{i_n,j_n\} \\
			{k_1,k_2}}}\pdv{k_2}\mel{i_0}{U^\dagger_{k_2}}{i_1}\pdv{k_1}\mel{j_1}{U_{k_1}}{j_0} \delta\left(k_1-k_2\right)\delta_{i_1,j_1}\label{sig2}
	\end{align}
	$\{i_n,j_n\} = 1,2$ and $k_1,k_2 \in \left[0,\pi\right)$.
	\begin{align}
		&\pdv{k_1}\sum_{j_0}\mel{j_1}{U_{k_1}}{j_0} = \left(-\iota 2J\sin(k_1)\right) \cross \bigg\lbrace \nonumber\\
		&\qquad \delta_{j_1,1}\left((t_1+t_2)\cos(V)e^{\iota 2J\left(t_1+t_2\right)\cos(k_1)}\text{\ }+\right.\nonumber\\
		&\left.\qquad\qquad\iota(t_1-t_2)\sin(V)e^{-\iota 2J\left(t_1-t_2\right)\cos(k_1)}\right)-\nonumber\\
		&\qquad\delta_{j_1,2}\left(-(t_1+t_2)\cos(V)e^{-\iota 2J\left(t_1+t_2\right)\cos(k_1)}\text{\ } + \right.\nonumber\\
		&\left.\qquad\qquad\iota(t_1-t_2)\sin(V)e^{\iota 2J\left(t_1-t_2\right)\cos(k_1)}\right)\bigg\rbrace
	\end{align}
	\begin{align}
		&\pdv{k_2}\sum_{i_0}\mel{i_0}{U^\dagger_{k_2}}{i_1} = \left(-\iota 2J\sin(k_2)\right) \cross \bigg\lbrace \nonumber\\
		&\qquad \delta_{i_1,1}\left(-(t_1+t_2)\cos(V)e^{-\iota 2J\left(t_1+t_2\right)\cos(k_2)}\text{\ }+\right.\nonumber\\
		&\left.\qquad\qquad\iota(t_1-t_2)\sin(V)e^{\iota 2J\left(t_1-t_2\right)\cos(k_2)}\right)+\nonumber\\
		&\qquad\delta_{i_1,2}\left((t_1+t_2)\cos(V)e^{\iota 2J\left(t_1+t_2\right)\cos(k_2)}\text{\ } - \right.\nonumber\\
		&\left.\qquad\qquad\iota(t_1-t_2)\sin(V)e^{-\iota 2J\left(t_1-t_2\right)\cos(k_2)}\right)\bigg\rbrace
	\end{align}
	Using these results, redefining $t_i\rightarrow t_i/\tau$ and some straightforward calculations, one arrives at
	\begin{equation}
	\begin{split}
	&\widetilde{\sigma}^2(t_1,t_2,V) =\frac{8}{L}\left(\sum_k \sin[2](k)\right)\cross\\&\qquad\left(J\tau\right)^2\left(\left(t_1+t_2\right)^2 - 4\sin[2](V)t_1t_2\right).
	\end{split}
	\end{equation}
	
	Replacing $\frac{1}{L}\sum_{k} \sin[2](k)\rightarrow \frac{1}{2\pi}\int_{0}^{\pi} \dd{k}\sin[2](k) = \frac{1}{4}$, we have 	\begin{equation}
	\widetilde{\sigma}^2(t_1,t_2,V) = 2\left(J\tau\right)^2\left(t_1^2+t_2^2 + 2 \cos(2V)t_1t_2\right)
	\end{equation}\\
	
	\section{Calculation of $Q(n;N\geq2)$}
	\label{appsec2}

	\subsection{General form of $Q(n;N\geq2$)}
	We begin by considering $Q(n;N=0)$, which is the probability that no kick has been imparted to the system until step $n$. From the definition,  
	\begin{align}
	Q(n;0) &= \sum_{n'=n+1}P_1(n')\\
&= 1 - \sum_{n'=1}^n P_1(n')
	\end{align}
	
	\begin{align}
	Q(n;1) &= \sum_{n'=1}^n\left(P_{1}(n') - P_{2}(n')\right)\nonumber\\
	&=1 - Q(n;0) - \sum_{n'=1}^nP_{2}(n')\nonumber\\
	\sum_{n'=1}^nP_{2}(n') &= 1 - Q(n;1) - Q(n;0)\nonumber\\
	Q(n;N\geq 2)&=\sum_{n'=1}^nP_{2}(n')
	\end{align}
	
	\subsection{$Q(n;N\geq 2)$ for Specific Cases}
	
	We first consider the case of $F(K) = U(K;A)$. We begin by calculating $P_1(n)$ and $P_2(n)$.
	
	The first kick can take place at any $1\leq n\leq A$, with equal probability.
	\begin{equation}
	P_1(n) = \begin{cases}
	\frac{1}{A},& 1\leq n\leq A\\
	0,& \text{otherwise}
	\end{cases}
	\end{equation}
	
	From Eq. \ref{PNdef}, $P_2(n) = P(K_0+K_1=n)$, with $\lbrace K_i\rbrace\leq A$. For $n\leq A+1$, there are $n-1$ ways of picking $\lbrace K_i\rbrace$, since choosing $K_0$ from 1 to $n-1$ decides $K_1$. For $n>A+1$, the upper limit on $\lbrace K_i\rbrace$ requires that $K_0$ can only take ($2A-n+1$) values from $n-A$ to $A$.
	\begin{equation}
	P_2(n) = \frac{1}{A^2}\begin{cases}
	n-1,& 2\leq n\leq A\\
	2A-n+1,& A< n\leq 2A\\
	0,& \text{otherwise}
	\end{cases}
	\end{equation}
	
	If $n\leq A$,
	\begin{align}
		Q(n;N\geq 2) &= \sum_{n'=1}^n P_{2}(n')\nonumber\\
		&=\frac{1}{A^2}\sum_{n'=2}^n (n'-1) \hspace{1em}\left(\text{by Eq.\ref{norml}}\right)\nonumber\\
		Q(n;1) &= \frac{n(n-1)}{2A^2}
	\end{align}
	
	If $n>A$
	\begin{align}
	Q(n;N\geq 2 ) &=\frac{1}{A^2}\sum_{n'=2}^{A} (n'-1)  + \frac{1}{A^2}\sum_{n'=A+1}^{n} (2A-n'+1)\nonumber\\
	&=\frac{2n - A-1}{2A} -\frac{(n-A-1)(n-A)}{2A^2}
        \label{QappB}
\end{align}

Putting these together,
	
	\begin{equation}
	Q(n;N\geq 2) = \begin{cases}
	\frac{n(n-1)}{2A^2},&n\leq A\\
	\frac{2n - A-1}{2A} - \frac{(n-A-1)(n-A)}{2A^2}, & A\leq n\leq 2A\\
	1,&n> 2A
	\end{cases}.
	\end{equation}
	
	Next, we consider $F(K) = P(K;\lambda)$, for which it is straightforward to calculate $P_N(n)$ for any $n$ and $N$. By construction, more than 1 kick at a time step is not allowed and thus we have
	\begin{equation}
	P(K;\lambda) \equiv e^{-\lambda}\frac{\lambda^{K-1}}{(K-1)!}\theta(K-1),
	\end{equation} where $\theta(x)$ is the Heaviside Step-Function with the convention that $\theta(0)=1$. Then, we get
	\begin{align}
		P_N(n) &=\sum_{\lbrace n_j\rbrace}{}^{'} \prod_{j=1}^NP(K=n_j;\lambda)
	\end{align}
	where prime indicates that the summation over $\lbrace n_j\rbrace$ is performed subject to the constraint $\sum\limits_{j=1}^N n_j = n$.
	\begin{align}
	   P_N(n)&=\theta(n-N)\sum_{\lbrace n_j\rbrace}{}^{'} e^{-N\lambda}\prod_{j=1}^N\frac{\lambda^{n_j-1}}{(n_j-1)!}\\
		&=\theta(n-N)e^{-N\lambda}\lambda^{n-N}\sum_{\lbrace n_j\rbrace}{}^{'} \frac{1}{(n_1-1)!\ldots(n_N-1)!} \nonumber
	\end{align} 
	Under the redefinition of $n_j-1\to n_j$, such that the condition under the sum becomes $\sum n_j = n-N$, and the multinomial expansion, we have
	
	\begin{equation}
	P_N(n) = \theta(n-N) e^{-N\lambda}\frac{(N\lambda)^{n-N}}{(n-N)!}
	\end{equation}

	The actual form of $Q(n;N\geq 2)$ is not particularly useful, and we require just the median of $P_2(n)$, which is known to be $\approx2\lambda$ for large $\lambda$.

\section{\label{calc3} $\sigma^2(N)$ for the 2-band coin-toss model}
\label{appsec3}
In the 2-band case, the quantity $\sigma^2$ has an exact solution in terms of a $4\cross4$ \lq\lq{}disorder matrix\rq\rq{}, introduced in \cite{PhysRevB.97.184308}. We begin by defining $\sigma^2(N)$ using the noisy operators,
	
	\begin{widetext}
	\vspace{-2.15em}
	\begin{align}
	&\sigma^2(N)=\sum_{\substack{\{i_n,j_n\} \\
			{k_1,k_2}}}\braket{\psi_0}{i_0}\braket{j_0}{\psi_0}\overline{\left( \prod_{n=1}^{N}\mel{i_{n-1}}{F^\dagger_{k_1}}{i_n}\mel{j_{n-1}}{F^T_{k_1}}{j_n}\right)} \frac{\partial^2}{\partial k_1\partial k_2}\delta\left(k_1-k_2\right)\\
	&\overline{\mel{i_{n-1}}{F^\dagger_{k_1}}{i_n}\mel{j_{n-1}}{F^T_{k_1}}{j_n}}=p_k\mel{i_{n-1}}{F^\dagger_{k_1}(\lambda)}{i_n}\mel{j_{n-1}}{F^T_{k_1}(\lambda)}{j_n}\nonumber\\
	&\hspace{13em}+ (1-p_k)\mel{i_{n-1}}{F^\dagger_{k_1}(0)}{i_n}\mel{j_{n-1}}{F^T_{k_1}(0)}{j_n}.
	\end{align}
	One can define the following matrices
	\begin{align}
	D_1&\coloneqq  p_k \left[F^\dagger_k(\lambda)\otimes F^T_k(\lambda)\right] + (1-p_k) \left[F^\dagger_k(0)\otimes F^T_k(0)\right]\nonumber\\
	D_2&\coloneqq  p_k \left[F^\dagger_k(\lambda)\otimes \partial_k\left(F^T_k(\lambda)\right)\right] + (1-p_k) \left[F^\dagger_k(0)\otimes \partial_k(F^T_k(0))\right]\nonumber\\
	D_3&\coloneqq  p_k \left[\partial_k(F^\dagger_k(\lambda))\otimes F^T_k(\lambda)\right] + (1-p_k) \left[\partial_k(F^\dagger_k(0))\otimes F^T_k(0)\right]\nonumber\\
	D_4&\coloneqq p_k\left[\partial_k(F^\dagger_k(\lambda))\otimes \partial_k(F^T_k(\lambda))\right] + (1-p_k) \left[\partial_k(F^\dagger_k(0))\otimes \partial_k(F^T_k(0))\right],
	\end{align}
	in terms of which one can express $\sigma^2(N)$ as
	\begin{align}
	&\sigma^2(N)=\nonumber\\
	&\frac{1}{L}\sum_{\substack{i_0,j_0\\i_N,j_N}}\sum_{n=1}^{N}\delta_{i_N,j_N}\left[ D_1^{n-1}D_4D_1^{N-n} + \sum_{m=1}^{n-1}\left(D_1^{m-1}D_3D_1^{n-m-1}D_2D_1^{N-n} + D_1^{m-1}D_2D_1^{n-m-1}D_3D_1^{N-n}\right)\right]_{\left( i_0,j_0,i_N,j_N\right)},
	\end{align}
	\end{widetext}
	where the $\delta_{i_N,j_N}$ follows from applying $\delta\left(k_1-k_2\right)$ in the discrete limit.	
	Certain recursion relations prove helpful in reducing the computational complexity of this evaluation to $\order{N}$. These are
	\begin{align}
	A(N)=A(N-1)D_1 + D_1^{N-1}D_4;\hspace{1em}A(0)\coloneqq 0\nonumber,\\
	B(N)=B(N-1)D_1 + \widetilde{B}(N-1)D_3; \hspace{1em}B(1)\coloneqq 0\nonumber,\\
	\widetilde{B}(N)=\widetilde{B}(N-1)D_1 + D_1^{N-1}D_2;\hspace{1em}\widetilde{B}(0)\coloneqq 0\nonumber\\
	C(N)=C(N-1)D_1 + \widetilde{C}(N-1)D_2; \hspace{1em}C(1)\coloneqq 0\nonumber\\
	\widetilde{C}(N)=\widetilde{C}(N-1)D_1 + D_1^{N-1}D_3;\hspace{1em}\widetilde{C}(0)\coloneqq 0\nonumber,\\
	\sigma^2(N)=\frac{1}{L}\sum_{i,j}\left[A(N)+B(N)+C(N)\right]_{i,j}.
	\end{align}
Note that the indices in brackets are not the true indices of the $4\times4$ matrix $D$. The recipe to switch between the two type of indices -- by employing the Kronecker Product -- is
	\begin{equation}
	\left[D_1^N\right]_{\left( i_0,j_0,i_N,j_N\right)}\coloneqq\left[D_1^N\right]_{2(i_0-1)+j_0, 2(i_N-1)+j_N}.
	\end{equation}

	\bibliography{biblio}
	\bibliographystyle{apsrev4-2}
\end{document}